\documentclass[10pt,twocolumn]{article}
\usepackage[top=1.9cm, bottom=1.9cm, left=1.8cm, right=1.8cm]{geometry}
\usepackage[sort&compress,numbers]{natbib}  %

\usepackage{times}  
\usepackage{helvet} 
\usepackage{courier}  
\usepackage[hyphens]{url}  
\usepackage{graphicx} 
\usepackage{subfigure}
\usepackage{natbib}  
\usepackage{caption} 
\usepackage{booktabs}
\usepackage{xcolor}

\usepackage[compact]{titlesec}
\titlespacing*{\section}{0pt}{*3}{3pt}
\titlespacing*{\subsection}{0pt}{*2}{2pt}
\usepackage{xspace}
\usepackage{flushend}

\newif\ifcomment
\commenttrue
\ifcomment
\newcommand{\sz}[1]{{\bf \textcolor{brown}{SZ: #1}}}

\newcommand{\cc}[1]{{\bf\textcolor{orange}{CC: #1}}}
\else
\newcommand{\sz}[1]{}
\newcommand{\cc}[1]{}
\fi

\title{``Learn the Facts About COVID-19'': Analyzing the Use of Warning Labels on TikTok Videos}

\begin{document}

\author{Chen Ling$^1$, Krishna P. Gummadi$^2$,  and Savvas Zannettou$^3$ $^4$\\[0.5ex]
\normalsize{$^1$Boston University, $^2$Max Planck Institute for Software Systems, $^3$TU Delft, $^4$Max Planck Institute for Informatics}\\
\normalsize ccling@bu.edu,  gummadi@mpi-sws.org, s.zannettou@tudelft.nl
}
\date{}

\maketitle

\begin{abstract}

During the COVID-19 pandemic, health-related misinformation and harmful content shared online had a significant adverse effect on society.
In an attempt to mitigate this adverse effect, mainstream social media platforms like Facebook, Twitter, and TikTok employed \emph{soft moderation interventions} (i.e., warning labels) on potentially harmful posts.
Such interventions aim to inform users about the post's content without removing it, hence easing the public's concerns about censorship and freedom of speech.
Despite the recent popularity of these moderation interventions, as a research community, we lack empirical analyses aiming to uncover how these warning labels are used in the wild, particularly during challenging times like the COVID-19 pandemic.

In this work, we analyze the use of warning labels on TikTok, focusing on COVID-19 videos.
First, we construct a set of 26 COVID-19 related hashtags, and then we collect 41K videos that include those hashtags in their description.
Second, we perform a quantitative analysis on the entire dataset to understand the use of warning labels on TikTok.
Then, we perform an in-depth qualitative study, using thematic analysis, on 222 COVID-19 related videos to assess the content and the connection between the content and the warning labels.
Our analysis shows that TikTok broadly applies warning labels on TikTok videos, likely based on hashtags included in the description (e.g., 99\% of the videos that contain \#coronavirus have warning labels).
More worrying is the addition of COVID-19 warning labels on videos where their actual content is not related to COVID-19 (23\% of the cases in a sample of 143 English videos that are not related to COVID-19).
Finally, our qualitative analysis on a sample of 222 videos shows that 7.7\% of the videos share misinformation/harmful content and do not include warning labels, 37.3\% share benign information and include warning labels, and that 35\% of the videos that share misinformation/harmful content (and need a warning label) are made for fun. 
Our study demonstrates the need to develop more accurate and precise soft moderation systems, especially on a platform like TikTok that is extremely popular among people of younger age.

\end{abstract}

\section{Introduction}
\label{sec:intro}

TikTok revolutionizes the world of user-generated videos by offering a platform that continuously recommends engaging short-length videos to users.
Nowadays, the platform is viral, in particular among young people, having over 1B users worldwide with a global penetration estimated at 18\% of all Internet users aged between 16 and 64~\cite{tiktok_statistics}.
TikTok is popular due to interesting and visually engaging videos focusing on various topics like dancing, comedy videos, do-it-yourself videos, etc.
At the same time, people use TikTok for malevolent purposes like spreading extremist videos~\cite{weimann2020research} or misinformation~\cite{basch2021global}, hence it is of paramount importance to have in place scalable and effective content moderation systems to mitigate negative effects from the spread of harmful content.

The need for timely and effective content moderation becomes more apparent when considering the spread of health-related misinformation. 
TikTok became extremely popular during the COVID-19 pandemic and, as with many other platforms on the Web (e.g., Facebook, Twitter, YouTube, etc.), the platform is tasked with dealing and mitigating the spread of COVID-19 related misinformation.
To facilitate content moderation, platforms use a combination of human moderators and automated means (e.g., Machine Learning classifiers) to remove or flag content that is deemed to be harmful or going against the platform's guidelines~\cite{grimmelmann2015virtues}.

Recently, platforms like Twitter, Facebook, and TikTok started opting for more ``softer'' moderation interventions in an attempt to inform users about the content posted on their platform without removing it~\cite{twitter_labels,facebook_labels,tiktok_labels}.
Soft moderation interventions help platforms ease users' concerns relating to censorship and suppression of free speech.
One of the most common and popular type of soft moderation interventions is the addition of warning labels on posts that are likely to be sharing harmful content.
Previous work focus on assessing the effectiveness of warning labels~\cite{bode2015related,kaiser2021adapting,moravec2020appealing}, how users interact with and perceive warning labels~\cite{mena2020cleaning,geeng2020social,saltz2020encounters,seo2019trust}, investigating unintended consequences that occur from warning labels~\cite{pennycook2020implied,nyhan2010corrections}, and analyzing how warning labels are used on platforms like Twitter~\cite{zannettou2021won,sanderson2021twitter}.
Despite this rich previous work on warning labels, we lack empirical analyses on the use of warning labels to tackle health-related misinformation, particularly on emerging platforms like TikTok.

In this paper, we perform a mixed-methods analysis of COVID-19 related TikTok videos and their respective warning labels (if any), intending to understand how they are applied on TikTok and analyze the content of the videos and whether they should get moderated.
To do this, we collect 41,853 TikTok videos that include COVID-19 related hashtags in their description, along with information on whether the videos have warning labels.
Then, we quantitatively analyze the dataset to gain a broad understanding of how TikTok applies warning labels on COVID-19 related videos.
Also, we perform an in-depth thematic qualitative analysis~\cite{tseng2020tools} in a random sample of 222 COVID-19 related videos to characterize their content and whether there is a justification for the inclusion/absence of warning labels (e.g., spreading misinformation or harmful content).
Our analysis yields several findings:

\begin{itemize}
  \item Our analysis shows that TikTok applies a broad and coarse-grained mechanism for moderating videos with warning labels, likely based on hashtags included in the videos' description.
  For instance, we find that 99\% of the videos in our dataset that contain \#coronavirus in the description have warning labels.
  \item More worryingly, we find a substantial percentage of videos that include COVID-19 warning labels on videos that their content is not related to the COVID-19 pandemic. In a sample of 500 randomly selected videos, we find 143 English videos unrelated to COVID-19, and 33 of them include COVID-19 warning labels (i.e., ``Learn the facts about COVID-19'' or ``Learn more about COVID-19 vaccines'').
  \item Our qualitative analysis (on a sample of 222 videos) shows that 37.3\% of the videos do not share misinformation or harmful content and they include a warning label, while 7.7\% of the videos share misinformation/harmful content and do not include a warning label.
  We also find that 35\% of the videos that share misinformation/harmful content and they need a warning label are made for fun.
\end{itemize}

\textbf{Implications.} Our findings have several implications for end-users and platforms:
\begin{itemize}
\item   The fact that warning labels are added to a large number of videos may cause users to ignore such warning labels when using the platform. In other words, if all videos include warning labels, it is doubtful that users will pay attention to them, hence reducing their utility.
  Overall, we argue that platforms should make a best-effort approach to show warning labels on potentially harmful content to maximize the utility and effectiveness of soft moderation interventions.
  Also, there is a need to use finer-grained warning labels based on the perceived risk from disseminating information included in videos.
  For instance, higher-risk videos (e.g., videos sharing confirmed misinformation) should have different warning labels than lower-risk videos (e.g., sharing unconfirmed information that will likely have little effect on the platform and the real world).
\item   The addition of COVID-19 warning labels to videos unrelated to COVID-19 might cause users to lose trust in moderation systems, as warning labels will be included in videos irrelevant to the target moderation context.
  Possible loss of trust in the platform's moderation system is likely to lead users to ignore such warning labels, diminishing warning labels' effectiveness.
\item The substantial percentage of videos that share misinformation or harmful content and do not include a warning label highlights the need to design and develop more precise and accurate content moderation systems, particularly on a platform like TikTok that is viral among young people.
\item Finally, the substantial percentage of fun videos that share misinformation/harmful content and require warning labels emphasize the need to raise awareness to end-users that even videos made for fun can share harmful information, which might be catastrophic when considering health-related information (e.g., COVID-19 information).
\end{itemize}

\section{Related Work}
\label{sec:relatedwork}

In this section, we provide an overview of the related work focusing on soft moderation interventions and the TikTok platform.

\textbf{Soft moderation interventions} refer to moderation actions performed on social media platforms and aim to inform users about the content (and its nature, e.g., it is likely to be misinformative) without removing the actual content.
The most popular soft moderation intervention is the addition of warning labels on user posts, an approach that is being used by mainstream social media platforms like Twitter, Facebook, and TikTok~\cite{twitter_labels,facebook_labels,tiktok_labels}.
These interventions aim to limit the spread of potentially harmful content on social media platforms by informing users, hence decreasing the likelihood of resharing harmful content.

According to Mena~\cite{mena2020cleaning}, warning labels assist in preventing the spread of deceptive content on social media on Facebook.
Another body of work investigates users' perceptions towards such interventions and their perceived effectiveness, finding that while the majority of users are positive about social media interventions, they usually resort to other ways to verify information (e.g., via web searches)~\cite{geeng2020social}.
Similarly,  Saltz et al.~\cite{saltz2020encounters} focus on users' perceptions, and they find that the use of warning labels is seen as excessively authoritarian, biased, and punitive.
Controlled experiments by Kaiser et al.~\cite{kaiser2021adapting} indicate that the design of warning labels can inform users about the risk of harm rather than moderating their behavior. 

Other previous work investigates various effects from the use of soft moderation interventions like the ``implied truth effect''~\cite{pennycook2020implied} (i.e., content without interventions are believed to be credible), the ``backfire effect''~\cite{nyhan2010corrections} (i.e., people strengthen their support for misinformation after seeing interventions), and the ``illusory truth effect''~\cite{pennycook2018prior} (people believe misinformation when exposed to it multiple times, despite the existence of interventions).

Finally, there is work focusing on empirical analyses of soft moderation interventions on the Web.
Zannettou~\cite{zannettou2021won} shows that tweets that had interventions on Twitter during the 2020 US elections received more engagement compared to tweets without interventions.
Similarly,  Sanderson et al.~\cite{sanderson2021twitter} focus on political content, specifically on tweets made by Donald Trump that received warning labels.
They find that the content of those tweets was also circulated on other platforms like Reddit, Facebook, and Instagram and that tweets with warning labels were posted more often on different platforms compared to tweets that did not have warning labels.

\textbf{TikTok.} TikTok is one of the platforms that started using soft moderation interventions, particularly warning labels, to limit the spread of harmful content.
According to their report, during the second quarter of 2021, TikTok attached warning labels on 1,856,773 videos that were viewed 11,169,599,491 times~\cite{tiktok2021q2}.
This report highlights the extensive use of warning labels on TikTok and motivates our work to study the use of warning labels on TikTok.

According to a recent review of social media's intended and unintended uses during the COVID-19 pandemic, social media platforms were critical avenues for understanding various phenomena like identifying infodemics, predicting COVID-19 cases, and analyzing government policies and people's reactions to policies~\cite{tsao2021social}.
Naturally, previous work focuses on TikTok to analyze users' behavior during the COVID-19 pandemic.
Ostrovsky and Chen~\cite{ostrovsky2020tiktok} analyze the top 100 TikTok videos with the hashtags ``COVID-19,'' ``COVID19,'' and ``coronavirus'' in July 2020, finding several videos shared by health care professionals as well as people documenting their experiences with the COVID-19 pandemic. They also find a small percentage (0.66\%) of videos sharing misleading or inaccurate information.
Similarly,  Basch et al.~\cite{basch2020covid} analyze 100 videos posted with the hashtag ``Coronavirus,'' finding that the most popular topic of discussion is anxiety caused by the COVID-19 pandemic and quarantine. They also analyzed videos shared by the World Health Organization, finding that they mostly share information about how the virus is transmitted, the symptoms, and sharing prevention tips.
Finally, other work focus on TikTok videos that share misinformation about masks~\cite{baumel2021dissemination} and how TikTok's virality is associated with disseminating public health messaging~\cite{eghtesadi2020facebook, southwick2021characterizing}.

\textbf{Remarks.} To the best of our knowledge, we perform the first empirical analysis on using soft moderation interventions (i.e., warning labels) on TikTok, particularly on COVID-19 related videos.
Our work is a stepping stone towards understanding these content moderation systems and subsequently helping in improving them.

\begin{table}
\centering
\resizebox{\columnwidth}{!}{
  \begin{tabular}{lr}
  \toprule
  \textbf{Warning Labels} & \textbf{\#Videos (\%)}\\ 
  \midrule
Learn more about COVID-19 vaccines & 12,372 (29.75\%)\\
Learn the facts about COVID-19 & 6,615 (15.90\%)\\
The action in this video could result in serious injury & 146 (0.35\%)\\
This content may not be suitable for some viewers & 3 (0.007\%)\\
\hline
\textbf{Total} & 19,136 (46.01\%)\\
\bottomrule
  \end{tabular}
  }
  \caption{Warning labels that exist in our dataset and the number/percentage of videos that include them.} 
  \label{tbl:types}
\end{table} 

\section{Dataset}
\label{sec:dataset}

TikTok allows users to search for videos with a specific hashtag and view the most popular results, as determined by TikTok's search algorithm.
We use TikTok's search feature to collect our dataset by scraping video search results using a set of COVID-19 related hashtags.
To construct the set of hashtags, we begin by scraping the videos from \#coronavirus and \#vaccine.
We start using these two specific hashtags mainly because of their popularity within the TikTok platform~\cite{coronatime}.
Then we iteratively expand our hashtag set by looking into the hashtags that co-appear in the descriptions of the collected videos.
In each iteration, for each hashtag, we extract the five most popular COVID-19 related hashtags that co-appear in the video descriptions (in terms of count) and then add them to our hashtag set.
To determine whether the hashtags are related to COVID-19, an author of this paper manually annotated each candidate hashtag as related to COVID-19 or not.
We repeat this iterative procedure two times on May 29, 2021, collecting a set of 26 COVID-19 related hashtags (available at~\cite{hashtags}) and 41,583 TikTok videos.

 For each TikTok video, we collect various metadata, including:
 (1) when the video was posted and by which user;
 (2) user engagement statistics (number of likes, comments, and reposts that the video received by the time of our data collection);
 (3) URL and description of the video;
 and (4) whether the video received any soft moderation intervention by TikTok (i.e., a warning label presented in TikTok's user interface).

\textbf{Ethical Considerations.} 
Our study only uses publicly available information, and there is no interaction with human subjects. 
Hence our work is not considered human subjects research by our institution's Ethical Review Board.
Nonetheless, there are important ethical considerations to be made when analyzing social media data.
Overall, we report results in aggregate, we do not attempt to deanonymize users, and we do not track users across sites~\cite{kenneally2012menlo}.

\begin{table}
\centering
\resizebox{\columnwidth}{!}{
\begin{tabular}{@{}lrlr@{}}
\toprule
\textbf{\begin{tabular}[c]{@{}l@{}}Hashtag\end{tabular}} & \textbf{\%Videos}           & \textbf{\begin{tabular}[c]{@{}l@{}}Hashtag\end{tabular}} & \multicolumn{1}{l}{\textbf{\%Videos}} \\ \midrule
\#coronavirus                                                                  & \multicolumn{1}{r|}{99.0\%} & \#mask                                                                           & 10.2\%                                \\
\#covidvaccine                                                                 & \multicolumn{1}{r|}{98.3\%} & \#quarantineroutine                                                              & 17.2\%                                \\
\#pfizervaccine                                                                & \multicolumn{1}{r|}{98.2\%} & \#quarantine                                                                     & 21.1\%                                \\
\#modernavaccine                                                                 & \multicolumn{1}{r|}{98.2\%} & \#mrna                                                                           & 22.8\%                                \\
\#covidvaccinesideeffects                                                        & \multicolumn{1}{r|}{97.7\%} & \#lockdown                                                                       & 23.1\%                                \\ \bottomrule
\end{tabular}%
}
  \caption{Top five hashtags with the larger/smaller percentage of videos that received  warning labels (e.g., 99\% of the videos that have \#coronavirus in their description received a warning label by TikTok).} 
  \label{tbl:percentage}
\end{table}

\begin{figure}[t]
\centering
\includegraphics[width=\columnwidth]{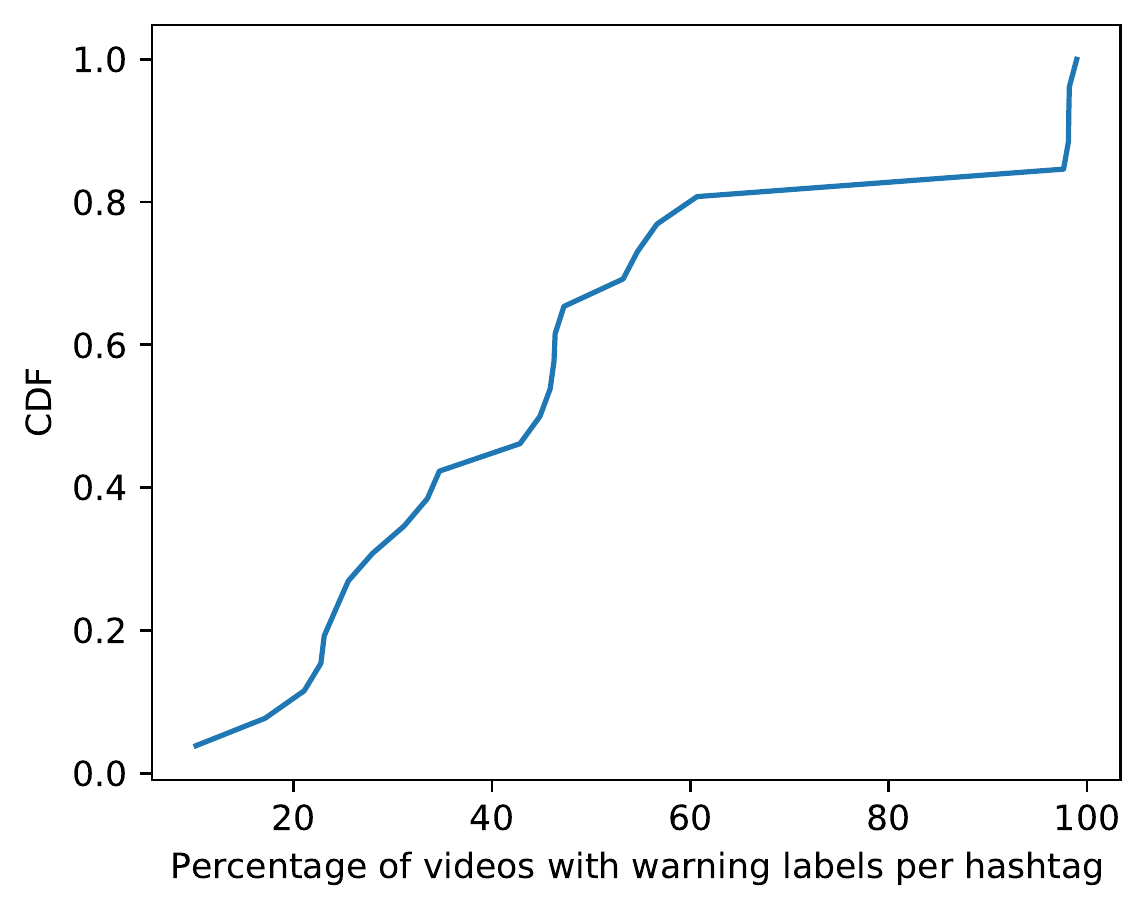}
\caption{CDF of the percentage of videos with warning labels per hashtag.}
\label{fig:hashtag}
\end{figure}

\begin{figure*}[t!]
\centering
\subfigure[]{\includegraphics[width=0.33\textwidth]{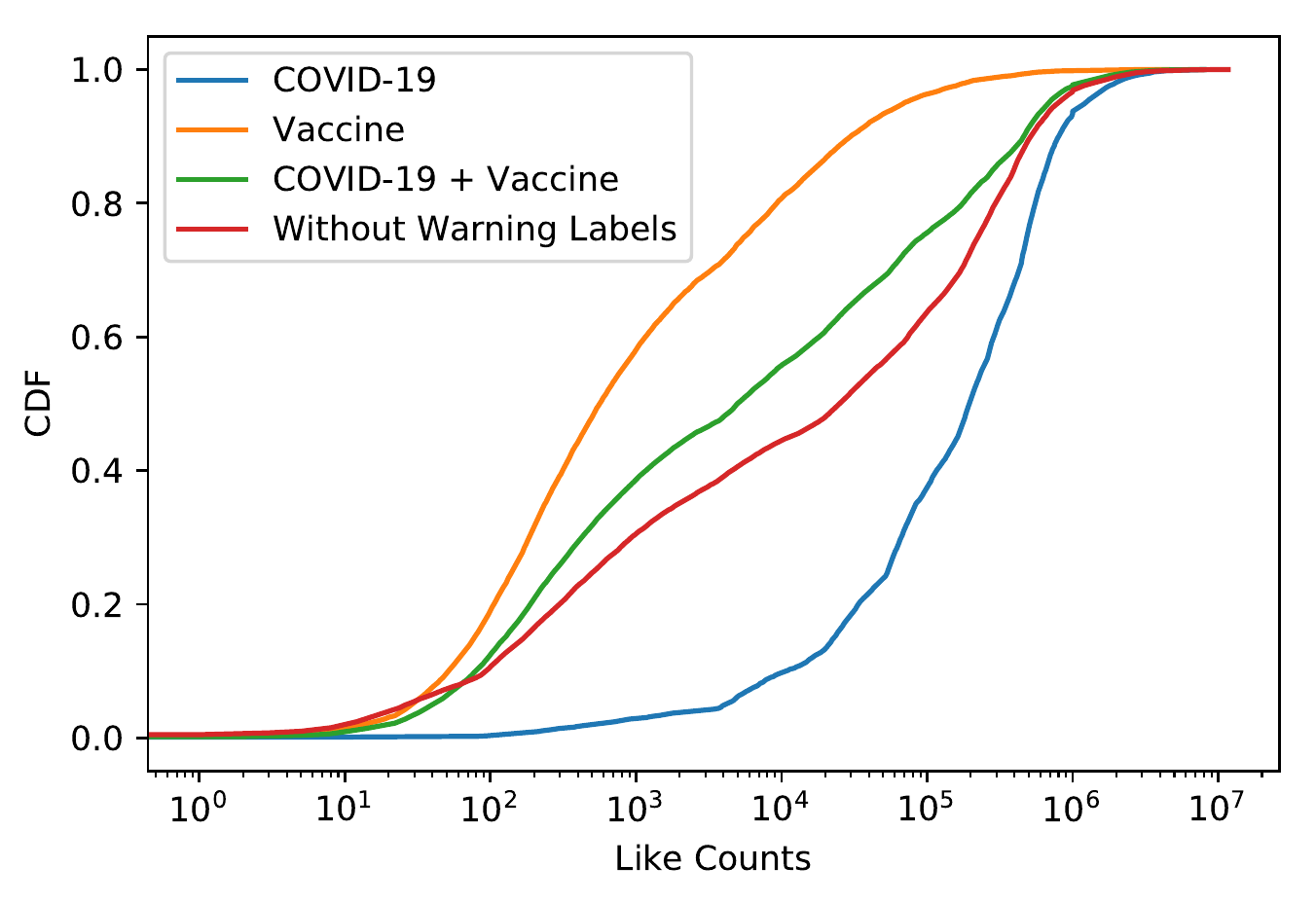}\label{fig:like}}
\subfigure[]{\includegraphics[width=0.33\textwidth]{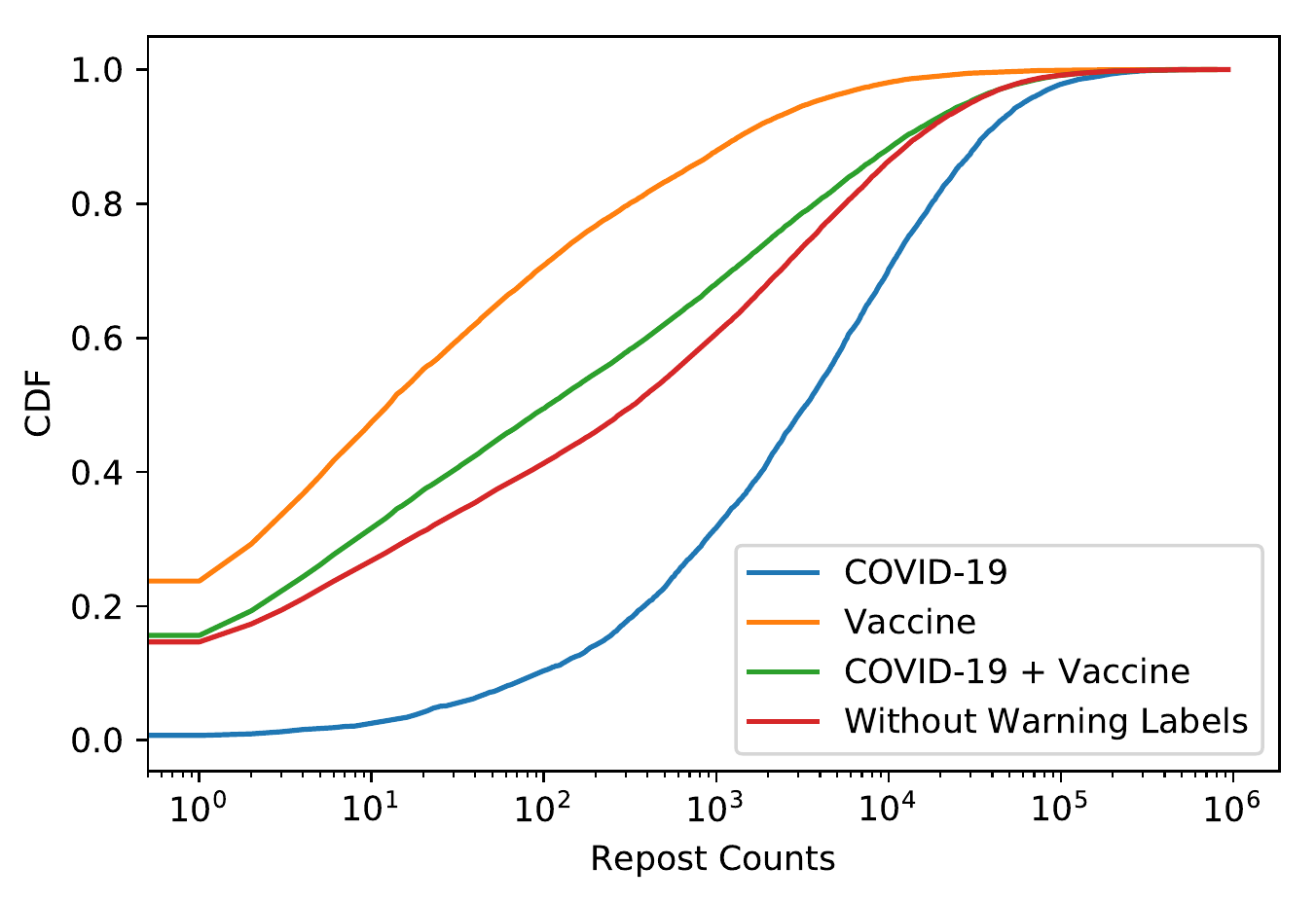}\label{fig:repost}}
\subfigure[]{\includegraphics[width=0.33\textwidth]{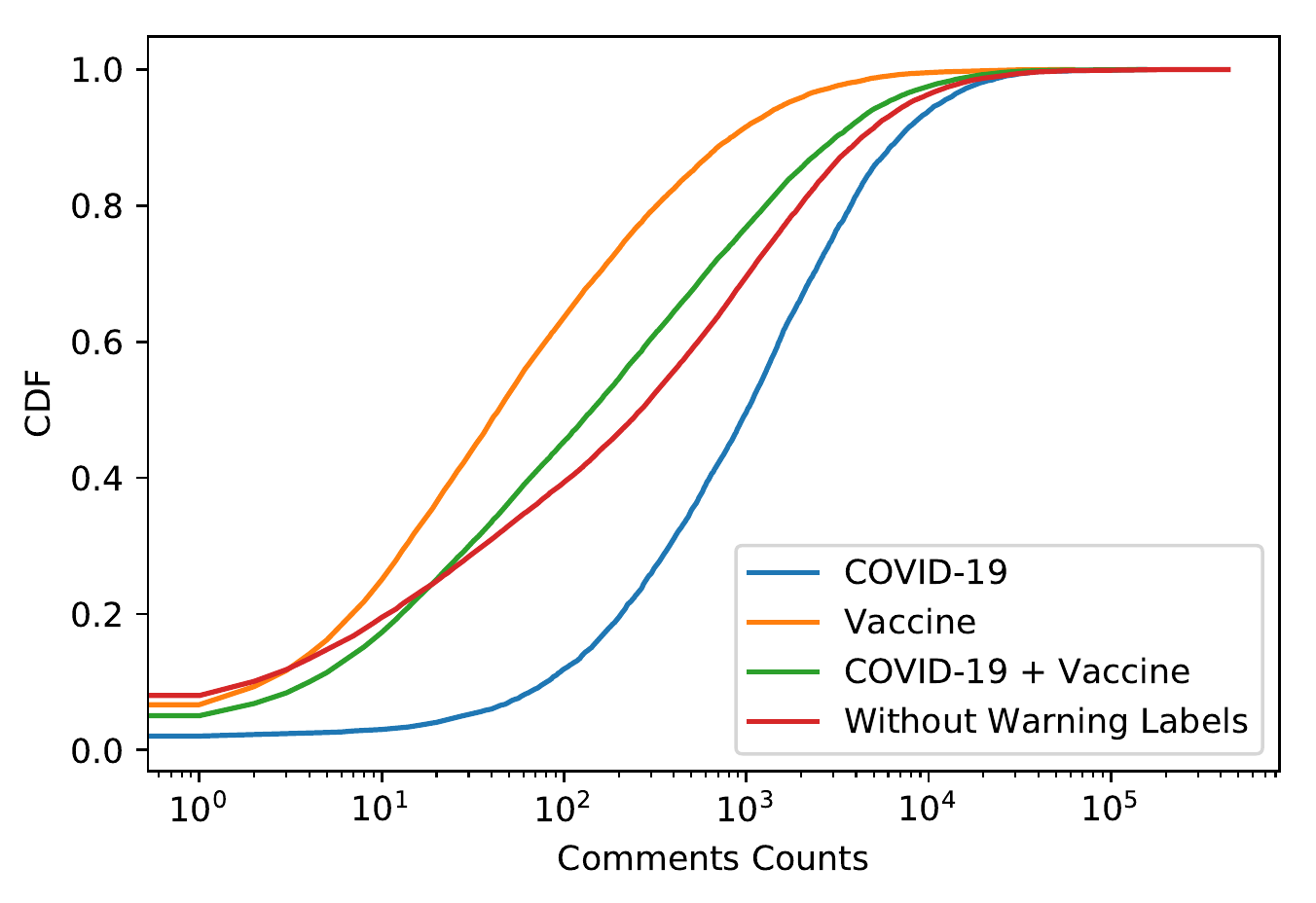}\label{fig:comments}}
\caption{CDF of the number of likes/reposts/comments for videos with and without warning labels.}
\label{fig:counts}
\end{figure*}

\section{Quantitative Analysis}
\label{sec:quantitative}

In this section, we perform an analysis on the entire dataset to quantitatively analyze the use of warning labels on TikTok.

\textbf{Prevalence of warning labels.} We start our analysis by looking into the prevalence of warning labels in our TikTok videos dataset.
Table~\ref{tbl:types} reports the number and percentage of videos that have a warning label in our TikTok dataset.
Overall, we find that 46\% of the videos in our dataset have a warning label, which is somewhat surprising since we expect only a small number of videos to be considered harmful and justify the inclusion of a warning label.
The most popular warning label is related to COVID-19 vaccines, with the warning label ``Learn more about COVID-19 vaccines'' appearing in 29\% of all the videos in our dataset.
The second most popular warning label is ``Learn the facts about COVID-19'' with 15.9\% of all the videos.
Both of these warning labels aim to inform users about the COVID-19 pandemic and the vaccines.
Also, we find a small percentage of videos that include warning labels unrelated to the COVID-19 pandemic, like ``The action in this video could result in serious injury'' (0.35\%) and ``This content may not be suitable for some viewers'' (0.007\%).
The appearance of these two warning labels likely indicates that some of the videos are unrelated to the COVID-19 pandemic, despite including COVID-19 related hashtags in their description.

Next, we look into the hashtags that exist in our dataset to shed some light on whether the addition of the warning labels is based on hashtags (i.e., all videos that contain a specific hashtag receive a warning label).
To do this, for each hashtag, we calculate the percentage of the videos that received a warning label over all the videos that include this specific hashtag.
Fig.~\ref{fig:hashtag} shows the cumulative distribution function (CDF) of the percentage of videos that received warning labels per hashtag, while Table~\ref{tbl:percentage} shows the five hashtags with the larger/smaller percentage of videos with warning labels.
We observe that five hashtags have a high percentage ($>97\%$) of videos that receive a warning label, namely, \#coronavirus, \#covidvaccine, \# pfizervaccine, \#modernavaccine, and \#covidvaccinesideeffects.
These hashtags refer to the COVID-19 pandemic and the vaccines; hence, it is likely that TikTok puts warning labels on almost all of the videos with these hashtags.
By manually looking into the small percentage of videos that do not include warning labels on these five hashtags, we do not find an apparent reason why these videos do not have warning labels.
When looking at the hashtags that had the smaller percentage of videos with warning labels, we find \#mask (10\%), \#quarantineroutine (17.2\%), \#quarantine (21.1\%), \#mrna (22.8\%), and \#lockdown (23.1\%).
These hashtags are also related to the COVID-19 pandemic, however, they received a substantially smaller percentage of warning labels.

\textbf{User Engagement.}
On TikTok, users can interact with videos by liking, reposting them, or commenting on them.
Here, we aim to analyze the user engagement on COVID-19 videos that receive warning labels and how this engagement compares to videos that did not.
To do this, we create three sets of videos:
1) Videos that receive the warning label ``Learn the facts about COVID-19'';
2) Videos that receive the warning label ``Learn more about COVID-19 vaccines'';
and 3) Videos that do not receive warning labels.
Fig.~\ref{fig:counts} shows the CDF of the number of likes, reposts, and comments made on videos of the three sets mentioned above, as well as a concatenation of the two first sets.
We observe that, in general, videos without warning labels receive more engagement than videos without warning labels across all metrics (median values of  24.8K vs. 4.9K for likes, 289 vs. 93 for reposts, and 245 vs. 132 for comments).
Our findings are in contrast with previous work focusing on Twitter during the 2020 US elections~\cite{zannettou2021won}, which show that tweets with politics-related warning labels received more engagement compared to tweets without warning labels.
This is likely because of differences in the moderation systems between TikTok and Twitter and because of the topic of discussion, as we expect people to be less polarized and less willing to engage with potentially harmful health-related content.

When looking into specific warning labels, we observe that the videos that received a COVID-19 warning label receive substantially larger engagement compared to the videos with the Vaccine warning label (median values of 194K vs. 559 for likes, 3.1K vs. 12 for reposts, and 967 vs. 42 for comments).
These results highlight that users are likely more hesitant to engage with videos that received warning labels about vaccines compared to videos that received warning labels about COVID-19.
Another possible explanation is that COVID-19 videos are more popular on TikTok in general compared to vaccine-related videos, however, we can not confirm this with the data we have.

\section{Qualitative Analysis}
\label{section:annotation}

While it is straightforward to automatically find videos that include hashtags related to COVID-19, there is a challenge in determining whether the content of the video is actually related to COVID-19.
This is because, usually, TikTok users put many hashtags in the video description, with some of them being unrelated to the video, likely in an attempt to attract more views/engagement.
For instance, \#foryou is one of the most popular hashtags on TikTok and is usually included in a large number of videos, irrespectively of the video's topic.
Due to this, simply selecting videos based on hashtags will yield many false positives (i.e., considering videos as COVID-19 related, when they are not related).
To overcome this challenge and construct a reliable dataset of COVID-19 videos, we extract a random sample of videos and we perform a manual annotation to determine whether videos are related to COVID-19 or not (Phase I).
Then, using the COVID-19 related videos in our sample, we develop a codebook that guides the thematic analysis of our TikTok sample dataset (Phase II).
Our thematic analysis helps us towards analyzing the content of these COVID-19 related TikTok videos with a particular focus on understanding the use of warning labels on such videos.
Below, we elaborate on the identification of COVID-19 related videos and the development of our codebook.

\subsection{Phase I: Identifying COVID-19 videos}

The first phase of our annotation process deals with identifying videos that are related to COVID-19.
Following the methodology from~\cite{tseng2020tools}, we randomly select 200 videos that contain a hashtag related to COVID-19 and have two authors of the paper review them and discuss them together to build a shared understanding of what a COVID-19 related video is.
Based on the initial annotation, we use the following definition for identifying COVID-19 related videos:
\textit{The video should express claims/opinions about the COVID-19 pandemic, sharing their life experiences about COVID-19, discussing or making fun of news related to the pandemic, or discussing policies/protective measurements that aim to limit the spread of the virus (e.g., wearing a mask, vaccinations, etc.).}

We then aim to test each author's ability to identify COVID-19 related videos independently.
To this end, we randomly choose 500 videos from our dataset, with the two authors labeling them as either related or not related to COVID-19.
During the annotation procedure, we exclude 129 videos not in English, and six videos that were unavailable (removed or deleted) by the time of the annotation.
For the rest of the videos, we find an almost perfect agreement between the two annotators by calculating Cohen's Kappa score~\cite{mchugh2012interrater} ($\kappa = 0.98$).
For the small number of videos with disagreement, the two annotators solved the disagreements by discussing the videos and reaching a conclusion on whether the video is related to COVID-19.
Overall, our annotation procedure indicates that a substantial number of videos with COVID-19 related hashtags are not related to COVID-19, as only 222 out of 365 videos (from 500 videos, we exclude the unavailable and non-English videos) are COVID-19 related. 
Interestingly, when looking at English videos unrelated to COVID-19 (143 videos), we find that 33 of them (23\%) have a COVID-19 related warning label.
This further compounds the findings from our quantitative section, highlighting the broad application of COVID-19 related warning labels, which even propagates to videos unrelated to COVID-19.

\subsection{Phase II: Characterizing COVID-19 videos}
While labeling videos as COVID-19 related is vital, it does little to characterize the actual content itself.
Hence, in this phase, we design a codebook that guides us towards understanding the content of COVID-19 videos and warning labels on TikTok.
To build our codebook and perform annotation, two authors of this paper independently watched the COVID-19 related videos (as determined by Phase I) and produced initial codes using thematic coding~\cite{braun2006using}. This step was a relatively loose process, aiming to help the annotators familiarize themselves with the content of such videos.
Then, the two annotators discussed these initial codes and went through multiple iterations, using a portion of the data to refine the codebook.
The process continued until the codebook reached stability, and additional iterations would not refine it further.
After several rounds of discussion between the two annotators, we derived three high-level properties relevant to COVID-19 videos: (1)~video types, (2)~stance towards various topics related to COVID-19, and (3)~suitability of a warning label (e.g., video sharing misinformation or is it harmful).
Below, we explain the three high-level properties in detail.

\textbf{Video types.} 
TikTok videos can have various types, including acting, animated infographic, documentary, news, oral speech, pictorial slideshow, and TikTok dance~\cite{li2021communicating}.
Therefore, the first property of our codebook is related to understanding the type of the video, particularly in the context of COVID-19 videos.
Based on our definition of the COVID-19 related videos and our dataset, we code the following types of a video:
\begin{itemize}
    \item \textbf{Personal news:} TikTok videos are becoming a popular way of recording and sharing personal updates.
    That is, people share their own experiences or personal news about the COVID-19 pandemic. E.g., people saying they got infected, people saying they are vaccinated, etc.
    \item \textbf{Made for fun:} These videos have a fun nature and try to entertain viewers (e.g., comedy videos). This type also includes videos that have a sarcastic nature aiming to entertain, while sharing information about COVID-19.
    \item \textbf{Informational video/Commenting news:} As the COVID-19 pandemic progressed, a considerable amount of news arose regarding the number of new cases and governments' policies for countering the spread of the virus.
    In our dataset, we find videos that try to explain various aspects of the COVID-19 pandemic or comment on news about the development of the pandemic (e.g., sharing the number of cases over time or trying to explain how masks/vaccines work, etc.) 
    \item \textbf{Opinion:} 
    This type refers to videos where the poster is sharing their opinion about the COVID-19 pandemic.
\end{itemize}

\textbf{Stances of topics.}
During the COVID-19 pandemic, governments enacted specific policies (e.g., social distancing) and advised people to follow specific hygiene measures (e.g., frequently washing their hands).
This likely influenced TikTok users to share videos related to these policies and hygiene measures, hence for each video, we code the following topics:

\begin{itemize}
    \item \textbf{Washing hands} is the most common way to protect people from getting infected, hence we code each video referring to washing hands with this topic.
    \item \textbf{Masks:} Another effective way to limit the spread of the COVID-19 virus is by wearing masks. Wearing masks was strongly advised or enforced by many governments during the pandemic. We code videos related to wearing masks with this topic (e.g., sharing opinions on masks, explaining how masks work, etc.)
    \item \textbf{Social distancing:} As communities reopen and people can move more often in public, they are advised to keep physical distance (usually 1.5m away) from other people, a measure known as social distancing. We code videos discussing or showing social distancing with this topic.
    \item \textbf{Quarantine:} Refers to staying home and away from others as much as possible to help prevent the spread of COVID-19. This measure was used during the pandemic, hence we code videos showing or discussing life during quarantine with this topic.
    \item \textbf{Vaccine:} Refers to the development and use of vaccines to stop the spread of the virus. Overall, there is a heated debate and a lot of misinformation about vaccines on social media, hence it is an important topic to study in TikTok videos. 
\end{itemize}

In particular, we code each topic according to the video's stance towards that specific topic: (1) \textbf{Anti stance}, the video shares content that opposes the proposed policy or protective measurement (e.g., anti-vaccine stance, anti-mask stance, etc.); (2) \textbf{Pro stance}, the video shares content that is positive towards the proposed policy or protective measurement (e.g., pro-vaccine stance, pro-mask stance, etc. ); and (3) \textbf{Not applicable}, the video does not show a clear stance towards the proposed policy or protective measurement.

\textbf{Suitability for a warning label.}
Online misinformation is a critical problem, which is further compounded when considering health information during the COVID-19 pandemic.
Motivated by this, social media platforms like Twitter, Facebook, and TikTok put soft moderation interventions (i.e., warning labels) on potentially harmful or misleading posts/videos.
Such interventions aim to inform the users about the nature of the post and provide information about the COVID-19 pandemic.
Since one of our goals is to understand the use of warning labels on TikTok, we code each video on whether it should include a warning label and the reason for requiring a warning label.
Specifically, we use the following codes:

\begin{itemize}
  \item \textbf{Misinformation:} The video includes false claims about the COVID-19 pandemic (e.g., vaccines cause autism, COVID-19 is caused by 5G, etc.). In such cases, we code the video as ``Misinformation'' and we argue that the video should have a warning label to inform the viewers.
  \item \textbf{Harmful behavior:} The video includes harmful behavior, e.g., promoting not wearing masks, intentionally coughing on other people, etc.
  \item \textbf{Other:} If the video contains offensive language or hateful content related to the COVID-19 pandemic (e.g., anti-Chinese sentiments due to COVID-19).
  \item \textbf{Not applicable:} The video's content does not justify the inclusion of a warning label.
\end{itemize}

To test the annotators' agreement on coding videos based on the codebook, we extract 50 randomly selected videos (out of the 222 COVID-19 related videos) and have each annotator independently code the videos.
Then, we measure the inter-annotator agreement, for each field in our codebook, using Cohen's Kappa score (see Table~\ref{tbl:agreement}).
Overall, we find substantial agreement (see~\cite{mchugh2012interrater} for interpretation of the Kappa scores) between the two annotators across all the fields with the warning labels' field having the least agreement ($k=0.673$), which indicates that annotating videos on whether they should include warning labels (e.g., misinformation or harmful content) is a subjective task.
Finally, to fully code the entire sample of the 222 COVID-19 videos, one of the annotators coded the rest of the videos.

\begin{table}
\centering
\small
  \begin{tabular}{lr}
  \toprule
  \textbf{Feature} & \textbf{Cohen's Kappa Score}\\
  \midrule
Video types & 0.789\\
Stance for hand washing & 0.789\\
Stance for masks & 0.771\\
Stance for social distance & 0.913\\
Stance for quarantine & 0.899\\
Stance for vaccine & 0.817\\
Suitable for warning labels &  0.673\\
\bottomrule
\end{tabular}
  \caption{Cohen's Kappa score for each field in our codebook (based on a random sample of 50 videos and two annotators).} 
  \label{tbl:agreement}
\end{table}

\subsection{Results}
\label{sec:qualitative}

In this section, we present the main findings from our qualitative analysis on 222 COVID-19 related videos.
We look deeper into the video content, including the type of the video and its stance towards various COVID-19-specific topics and whether videos should include warning labels.

\subsubsection{Video types}
We start by looking into the various video types and whether videos for each kind should include a warning label.
Table~\ref{tbl:videotype} reports the numbers of videos in each video type and code for warning labels.
Overall, we find that 40\% of the videos share personal news/updates, 23\% of the videos are made for fun, 27\% are informational videos, while 10\% of the videos are opinion videos.

Over the entire dataset, we find 37 videos (16.6\%) that need a warning label, with most of them annotated as misinformation (73.4\% of videos that require warning labels).
Interestingly, we observe that ten videos are made for fun, and five videos share personal news. However, they also include misinformation claims related to COVID-19, hence a warning label should accompany these videos.
We also find nine videos that aim to be informational (i.e., inform viewers about a specific aspect of COVID-19), however, at the same time, they are sharing false claims.
Overall, these findings indicate that misinformation and harmful behavior can be promoted on videos that are seemingly personal or are made for fun, which has to be taken into account by moderation systems/human moderators, especially when it comes to health-related matters like COVID-19.

We present an example of a video made for fun and should include a warning label (TikTok also applied a warning label).
The video shows an anti-quarantine stance and contains harmful behavior relating to the COVID-19 pandemic and the spread of the virus.
The video was shared on March 12, 2020, when the first lockdown went into place, and it shows two boys at home partying during the quarantine.
At some point, they start quarreling and fighting loudly, and then one of the boys spits three times towards the other boy in an attempt to ``spread coronavirus.''
The uploader also added text to the video, saying, ``Why coronavirus is spreading so quickly.''
Overall, the video is for entertainment purposes, however, it demonstrates potentially harmful behavior in spreading the COVID-19 virus.
TikTok labeled this video with the warning label ``Learn the facts about COVID-19.''

\begin{table}
\centering
\resizebox{\columnwidth}{!}{
\begin{tabular}{@{}lrrrr@{}}
\toprule
\textbf{}        & \multicolumn{1}{c}{\textbf{\begin{tabular}[c]{@{}c@{}}Personal \\ News\end{tabular}}} & \multicolumn{1}{c}{\textbf{\begin{tabular}[c]{@{}c@{}}Made \\ for Fun\end{tabular}}} & \multicolumn{1}{c}{\textbf{Informational}} & \multicolumn{1}{c}{\textbf{Opinion}} \\ \midrule
\textbf{Not applicable}   & 78                                                                                    & 38                                                                                   & 50                                         & 19                                   \\
\textbf{Misinformation}   & 5                                                                                     & 10                                                                                   & 9                                          & 3                                    \\
\textbf{Harmful behavior} & 5                                                                                     & 2                                                                                    & 0                                          & 0                                    \\
\textbf{Other}            & 1                                                                                     & 1                                                                                    & 1                                          & 0                                    \\ \midrule
\textbf{Total}            & 89                                                                                    & 51                                                                                   & 60                                         & 22                                   \\ \bottomrule
\end{tabular}%
}
  \caption{Video types and suitability for a warning label.} 
  \label{tbl:videotype}
\end{table}

\subsubsection{Stance of topics}

Next, we look into the various topics that are included in the sampled dataset and the video's stance towards them. Table~\ref{tbl:videotopics} reports the number of videos in each topic and stance.

\textbf{Vaccine is the most discussed topic.} There are 222 COVID-19 related videos in our TikTok dataset, and a video may have stance of multiple topics, e.g., talking about masks and washing hands, or may not have a stance at all, e.g., not mentioning masks, or vaccine, etc.
In our sampled dataset, the most discussed topic is ``vaccine,'' with 46.3\% of the videos talking about them (15 videos have an anti-vaccine stance, while 65 videos have a pro-vaccine stance).
Other popular topics are ``masks,'' with 22.5\% of the videos, ``quarantine'' with 15.0\% of the videos,  ``social distance'' with 12.2\% of the videos, and 4\% of the videos are about ``washing hands.''

Here, we present an example of a video about vaccines, which should include a warning label, however, TikTok did not apply a warning label.
The video was shared on May 2, 2021, and depicts a phenomenon known as the Vaccine Gang~\cite{vaccine_gang}.
The video poster interacts with other users' videos in which they take vaccines other than Moderna to brag about how wonderful it is to take the Moderna vaccine.
The creator attacks people who use Pfizer because they need to renew their annual subscription to take more shots. In contrast, people who take Moderna do not need to wear glasses after being vaccinated (the Moderna vaccine improves eyesight).
Though it is an entertaining video, it shares false or unconfirmed information about vaccines.

\begin{table}
\centering
\resizebox{\columnwidth}{!}{
\begin{tabular}{@{}lrrrrr@{}}
\toprule
\textbf{}   & \multicolumn{1}{c}{\textbf{Quarantine}} & \multicolumn{1}{c}{\textbf{\begin{tabular}[c]{@{}c@{}}Washing\\hands\end{tabular}}} & \multicolumn{1}{c}{\textbf{Masks}} & \multicolumn{1}{c}{\textbf{\begin{tabular}[c]{@{}c@{}}Social\\  distance\end{tabular}}} & \multicolumn{1}{c}{\textbf{Vaccine}} \\ \midrule
\textbf{No stance}   & 196                                     & 215                                                                                   & 183                                & 201                                                                                     & 142                                  \\
\textbf{Anti stance} & 2                                       & 0                                                                                     & 5                                  & 2                                                                                       & 15                                   \\
\textbf{Pro stance}  & 24                                      & 7                                                                                     & 34                                 & 19                                                                                      & 65                                   \\ \bottomrule
\end{tabular}%
}
\caption{Topics of videos and their stance.}
\label{tbl:videotopics}
\end{table} 

\textbf{Vaccine is the most controversial/misinformative topic.} Out of 80 vaccine videos, 81.3\% videos are pro-vaccine, while 18.7\% videos are anti-vaccine.
The percentage of the videos with an anti-stance for vaccines is higher than other topics, hence it is the most controversial topic in our sample.
Specifically, out of 39 masks-related videos, 87.2\% videos are pro-mask, and 12.8\% videos are anti-mask.
Out of 26 quarantine videos, 92.4\%  videos are pro-quarantine, and only 7.6\%videos are anti-quarantine.
Out of 21 social distance videos, 90.4\%videos are pro-social-distance, and 9.6\%videos are anti-social-distance.
Finally, for washing hands, all seven videos advocate washing hands (pro-stance).
Overall, these findings indicate that most of the government's protective measurements and policies are discussed on TikTok, with some creators sharing their anti-stance ideology, mainly when talking about COVID-19 vaccines.

\begin{table}
\centering
\resizebox{\columnwidth}{!}{
\begin{tabular}{@{}lrrrrr@{}}
\toprule
\textbf{}        & \multicolumn{1}{c}{\textbf{Quarantine}} & \multicolumn{1}{c}{\textbf{\begin{tabular}[c]{@{}c@{}}Washing \\ hands\end{tabular}}} & \multicolumn{1}{c}{\textbf{Masks}} & \multicolumn{1}{c}{\textbf{\begin{tabular}[c]{@{}c@{}}Social\\ distance\end{tabular}}} & \multicolumn{1}{c}{\textbf{Vaccine}} \\ \midrule
\textbf{Misinformation}   & 0                                       & 0                                                                                     & 2                                  & 0                                                                                      & 15                                   \\
\textbf{Harmful behavior} & 2                                       & 0                                                                                     & 3                                  & 2                                                                                      & 0                                    \\ \bottomrule
\end{tabular}
}
  \caption{Topics of videos that have an anti stances and the respective warning label types.} 
  \label{tbl:antitype}
\end{table} 

We coded 16.7\% (37 out of 222) videos in our sample that have to include a warning label; 64.8\% (24 out of 37) of these labels are put to the videos that present an anti stance to the COVID-19 measurements (see Table~\ref{tbl:antitype}).
Specifically, all anti-vaccine videos in our dataset share some false claims about the effectiveness or side effects of vaccines, hence we argue that such videos should include warning labels.
Interestingly, we also find videos with an anti-stance on other safety measurements and promote harmful behavior (i.e., not following the guidelines aiming to stop the spread of COVID-19).
We find three that are anti-mask and promote harmful information, two that are anti-social-distance, and two that are anti-quarantine.

\textbf{Positive stance can be misinformative.}
We look in more depth into the videos that have a positive stance towards various topics (see Table~\ref{tbl:protype}).
Overall, we find a small number of videos that have a pro-stance towards masks (three videos), social distance (three videos), and vaccines (two videos) that are sharing either misinformation or promoting harmful behavior, hence they should include a warning label.
An example demonstrating how a video with a pro stance can be misinformative is about a Jamaican COVID-19 remedy.
This video was posted on February 1, 2021, and TikTok did not include a warning label on this video.
The video shows two women wearing masks (pro-mask) preparing a Jamaican remedy to cure them of COVID-19.
They barbecue an orange on the stove until the outer part of the orange turns black. Then, they peel the skin, mash the orange, and put some brown sugar in it.
Then, they start eating the orange, claiming that this remedy will cure COVID-19 and bring their taste and smell back.
This video shares false information about COVID-19 cures. We argue that this video should include a warning label to inform the viewers that the information/remedy included in this video is inaccurate.

\begin{table}
\centering
\resizebox{\columnwidth}{!}{
\begin{tabular}{@{}lrrrrr@{}}
\toprule
\textbf{}                 & \multicolumn{1}{c}{\textbf{Quarantine}} & \multicolumn{1}{c}{\textbf{\begin{tabular}[c]{@{}c@{}}Washing\\ hands\end{tabular}}} & \multicolumn{1}{c}{\textbf{Masks}} & \multicolumn{1}{c}{\textbf{\begin{tabular}[c]{@{}c@{}}Social\\ distance\end{tabular}}} & \multicolumn{1}{c}{\textbf{Vaccine}} \\ \midrule
\textbf{Misinformation}   & 0                                       & 0                                                                                    & 2                                  & 1                                                                                      & 2                                    \\
\textbf{Harmful behavior} & 0                                       & 0                                                                                    & 1                                  & 2                                                                                      & 0                                    \\ \bottomrule
\end{tabular}
}
  \caption{Topics of videos that have a pro stances and the respective warning label types.} 
  \label{tbl:protype}

\end{table}

\subsubsection{False placement of warning labels}

Our quantitative analysis (see Section~\ref{sec:quantitative}) shows that almost half of videos in our dataset are moderated with a warning label, and 
for specific hashtags (e.g., \#coronavirus, \#covidvaccine, etc.), $>97\%$ of videos have warning labels.
Here, we aim to assess how accurate is TikTok's moderation system with respect to adding warning labels on potentially harmful videos.
We do this by qualitatively analyzing the content of the video to identify misinformation, harmful content, or other reasons for the inclusion of a warning label.
Note that we assume that only videos that include misinformation or other harmful information should receive a warning label. 

\textbf{False positives.} We find 37.3\% (83 out of 222) false positive videos in our sample (i.e., the video should not include a warning label because it does not share misinformation/harmful content, and TikTok applied one of the two warning labels related to COVID-19).
Looking into the types of false positive videos, we find that the false positives are distributed across the various types of videos; 36.1\% of the false positives are ``Personal News,'' 24.1\% are made for fun, 30.1\% are information videos and 9.7\% are opinion videos.
When looking at the topics of the false positives, we find that the vaccine topic is the most popular, with 50.7\% of the false positives in our sample.
Other popular topics include masks (19.4\%), quarantine (11.9\%), social distance(10.4\%), and washing hands(7.6\%).
Interestingly, our annotations indicate that all false positive videos in our sample hold a pro-stance towards their respective topics.

Here, we present an example of a false positive video shared by the official TikTok account of the World Health Organization on March 21, 2020.
The video aims to raise awareness to people about the pandemic and highlights that young people can get infected by COVID-19.
Also, it mentions that young people might require treatment in hospitals when infected from COVID-19.
Overall, the video does not share any misinformation or harmful information, hence our annotation concludes that the video should not have a warning label.
On the other hand, TikTok added the "Learn the facts about COVID-19" warning label on this video.

\textbf{False negatives} We also find 7.7\% (17 out of 222) false-negative videos in our sample (i.e., the video should include a warning label because it shares misinformation/harmful content, and TikTok did not apply one of the two warning labels related to COVID-19).
The most frequent type of video in the false negatives is informational (35.2\%), then personal news (29.5\% ), then videos made for fun (29.5\%), and opinion videos (5.8\%).
Similar to the result of our false positive, the most popular topic in false negatives is the vaccine, with 47\% of all false positives.
Looking at the stances, we find that most false-negative videos have an anti-stance; 64.6\% of the false positives have an anti-stance to any topic.

Finally, we present an example of a false negative video shared on March 18, 2021.
The video has a text saying ``When you take the covid-19 vaccine'' and shows a woman sitting on a couch.
The woman starts having fast and intense spasms that cause her to jump on and off the couch.
Overall, this video is likely shared for entertainment purposes, but it shares misinformation about vaccine side effects (i.e., vaccines do not cause spasms).
TikTok did not apply any warning label on this video, however, we argue that this video should have a warning label that will inform viewers about the confirmed vaccine side effects.

\section{Discussion \& Conclusion}
\label{sec:discussion}

In this paper, we analyzed TikTok's use of warning labels on COVID-19 related videos.
We created a set of 26 COVID-19 related hashtags, and then we collected metadata for 41K videos that emerge after searching for the hashtag on TikTok.
We performed a mixed-methods analysis to understand the use of warning labels on TikTok, particularly on COVID-19 related videos.
Specifically, we performed a quantitative analysis on the entire dataset to analyze the prevalence of warning labels in our dataset, the interplay between the hashtags and the warning labels and analyze the engagement of video with and without warning labels.
Then, we built a codebook and performed a qualitative analysis that enabled us to analyze the video's content and better understand the connection of content and warning labels.
Below, we discuss our main findings and their implications for various stakeholders involved in online content moderation.

\textbf{Broad placement of warning labels.} Our quantitative analysis shows that TikTok broadly applies warning labels, likely by considering hashtags included in the video's description.
For instance, we find that 99\% of the videos that include \#coronavirus in our dataset are soft moderated with COVID-19 related warning labels.
At the same time, TikTok's COVID-19 warning labels are so generic that they cannot differentiate between high and low-level risks (i.e., videos that can have a disproportionately negative impact on society like dangerous false remedies about COVID-19 vs. videos that mention COVID-19).
Both the broad and generic nature of the application of warning labels have essential implications for end-users as they can compromise the effectiveness of warning labels.
This is because users are likely to ignore warning labels that appear in almost all videos, particularly in cases where the warning labels are very generic and unrelated to the video.
Indeed, previous studies on consumer warning labels (i.e., warning labels added on products like cigarettes)~\cite{robinson2016consumer} indicates that broad and generic use of warning labels can lead to people ignoring the warning labels altogether.
Overall, our findings highlight the need to develop fine-grained soft moderation systems that can provide accurate and specific warning labels to allow end-users to assess the risks of the information shared online.
The need for such moderation systems becomes pressing, particularly when considering platforms like TikTok, which is popular among young people, and health-related information (e.g., COVID-19 virus) that can have devastating effects on society (e.g., people dying because of misinformation).

\textbf{Substantial percentage of false positives/false negatives.} Our qualitative analysis on a sample of 222 COVID-19 related videos shows that 37\% of the videos that include warning labels do not share misinformation/harmful content, while at the same time 7.7\% of the videos that do share misinformation/harmful content have no warning labels.
This indicates that the use of warning labels by TikTok is not ideal and it is likely to hinder their utility and effectiveness.
In particular, the inaccurate use of warning labels can cause end-users to lose their trust in the platform and its content moderation system.
Subsequently, users can elect to either stop using the platform or completely ignore the warning labels, which means that warning labels can either ``backfire'' (i.e., platform losing users) or have limited use/effectiveness.
These findings call attention to the need for greater transparency by the platforms on how these warning labels are applied to online content.
By doing so, end-users will better understand why they see warning labels on specific videos and help them trust moderation systems~\cite{naher2019improving}. At the same time, it will help the research community better understand these systems and potentially improve them.

\textbf{Made for fun videos sharing misinformation.} Our qualitative analysis shows that videos made for fun can be misinformative and should be moderated with warning labels.
This finding further compounds previous work on Twitter, finding that racism and hate speech can be disseminated through the use of humor, sarcasm, and irony~\cite{petray2017your}.
The use of entertainment videos for spreading harmful content is a worrying trend and is likely indicating that, by using platforms like TikTok, we are ``amusing ourselves to death''~\cite{postman2006amusing}.
Overall, this finding emphasizes the need to raise end-user awareness that entertainment and comedy videos can share harmful content or misinformation.
A way to achieve this can be via more precise and detailed warning labels that inform end-users that the information included can be harmful despite the funny nature of the video.

\textbf{Limitations.} We conclude with the limitations of our work.
First, our data collection and the dataset are biased towards popular content. This is because we use TikTok's search functionality, which returns a limited number of popular videos according to their proprietary and closed algorithm.
Since TikTok does not provide any Application Programming Interfaces to collect large-scale datasets from the platform, we cannot quantify the bias in our dataset.
Nevertheless, we argue that for this study, the collected dataset, despite its biases, can shed light on how warning labels are used on TikTok.
Second, our qualitative analysis is based on coding a small number of videos (222), focusing only on English-speaking videos related to COVID-19.
Due to this limitation, we cannot assess the use of warning labels across multiple languages/countries.

\bibliographystyle{abbrv}


\end{document}